\documentclass{article}
\usepackage{amsfonts}
\usepackage{graphicx}

\evensidemargin=\oddsidemargin
\def\Title#1#2#3{%
    \baselineskip=18pt
    \begin{center}
          {\large\bf{#1} \\ }
          \bigskip\bigskip
          {#2} \\
          {#3} \\
    \end{center}}
\long\def\Abstract#1{%
         \bigskip
         \parbox{0.93\textwidth}{%
                 \begin{center}
                       {\bf Abstract} \\
                 \end{center}
                 \medskip{\baselineskip=14pt #1}
                 \vss}
         \bigskip}

\makeatletter
\renewcommand{\section}%
 {\@startsection{section}{1}{0pt}%
  {-3.25ex plus -1ex minus -.2ex}{1.5ex plus .2ex}%
  {\vspace*{5mm}\raggedright\large\bf }}
\renewcommand{\subsection}%
 {\@startsection{subsection}{2}{0pt}%
  {-2.25ex plus -.5ex minus -.2ex}{-1.5ex plus -.2ex}{\bf }}
\renewcommand{\subsubsection}%
 {\@startsection{subsubsection}{3}{0pt}%
  {-1.25ex plus -.2ex minus -.1ex}{-1.2ex plus -.2ex}{\bf }}

\makeatother

\begin{document}

\Title{The extended phase space approach\\
to quantization of gravity and its perspective}
{T. P. Shestakova}%
{Department of Theoretical and Computational Physics,
Southern Federal University,\\
Sorge St. 5, Rostov-on-Don 344090, Russia \\
E-mail: {\tt shestakova@sfedu.ru}}

\Abstract{The prerequisites of the extended phase space approach to quantization of gravity, which is alternative to the Wheeler -- DeWitt one and other existing approaches, are presented. The features of the proposed approach and conclusions from its underlying ideas are discussed.}

\section{Introduction}
I shall present the extended phase space approach to quantization of gravity \cite{SSV1,SSV2,SSV3,SSV4,Shest1}. The approach is alternative to existing approaches to quantization of gravity. The main idea, which it is based upon, is:

{\bf The quantization of gravity implies consideration of spacetimes with a nontrivial topology. In this case, the gravitating system has no asymptotic states, and this fact distinguishes gravity from other gauge fields.}

The founders of quantum geometrodynamics, the first approach to quantization of gravity, like Wheeler, Hawking and others, spoke that the Universe may have a nontrivial topology. However, this conjecture on a nontrivial topology appears to be in contradiction with the assumption on asymptotic states that is used in the path integral quantization of gauge theories.

To understand this contradiction, let us start with a brief review of quantization methods. The most of approaches to quantization of gravity have been elaborated to be consistent with the Dirac quantization scheme for constrained theories. Dirac remembered that he had been excited by the role that the Hamiltonian formalism had played when quantum mechanics had been created. He wrote in his ``Lectures on quantum mechanics'' \cite{Dirac1} that
\begin{quote}
``\ldots if we can put the classical theory into the Hamiltonian form, then we can always apply certain standard rules so as to get a first approximation to a quantum theory.''
\end{quote}

As it is known, Dirac faced the problem, how to construct the Hamiltonian formalism for a constrained system, since in this case, the Hamilton function cannot be constructed by the usual rule,
\begin{equation}
\label{Ham_func}
H=p_a\dot q^a+\pi_{\alpha}\dot\lambda^{\alpha}-L.
\end{equation}
Here, all generalized coordinates are divided into two groups: $\{q^a\}$ are the so-called physical variables, while $\{\lambda^{\alpha}\}$ are gauge, or non-physical degrees of freedom, the velocities of which cannot be expressed from the momenta equations,
\begin{equation}
\label{mom}
\pi_{\alpha}=\frac{\partial L}{\partial\dot\lambda^{\alpha}}=0
\end{equation}

It is worth noting that, in the beginning, Dirac included gauge variables into phase space together with physical ones, otherwise he would not get the secondary constraints by means of the Poisson brackets,
\begin{equation}
\label{P.B.}
\dot\pi_{\alpha}=\{\pi_{\alpha},H\}=\varphi_\alpha,
\end{equation}
but afterwards he declared that they are not of physical interest, redundant degrees of freedom.

Dirac is believed to find the solution to the problem by introducing the two postulates:
\begin{itemize}
\item One should add a linear combination of constraints $\{\varphi_{\alpha}\}$ to the Hamiltonian:
\begin{equation}
\label{Dir_Ham}
H=H_0+\lambda^{\alpha}\varphi_{\alpha}.
\end{equation}
\item When quantizing, the constraints in the operator form become conditions imposed on the state vector:
\begin{equation}
\label{quant_constr}
\varphi_{\alpha}|\Psi\rangle=0.
\end{equation}
\end{itemize}

Why these rules are postulates? They cannot be derived from other fundamental physical statements, also, they cannot be justified by a reference to the correspondence principle. Moreover, these postulates have never been verified by any physical experiments, while very successful theories, confirmed experimentally, are based on different methods.

For example, quantum electrodynamics is based on the Lagrangian formalism and perturbation theory. Ironically, the Dirac approach is used only in various attempts to quantize gravity, in other words, in the sphere where, until now, we have not got any experimental data.

Meanwhile, the development of quantization methods gave a hint how Hamiltonian dynamics can be constructed differently. In the path integral quantization of gauge theories, the gauge-invariant action of an original theory is replaced by an effective action which includes gauge fixing and ghost terms. A gauge condition can be chosen in such a way that it would introduce missing velocities into the effective Lagrangian. An example is given by the Lorentz gauge in electrodynamics,
\begin{equation}
\label{eff_act}
S_{ED}\rightarrow S_{eff}=\int d^4x\left({\cal L}_{ED}+{\cal L}_{gf}+{\cal L}_{ghost}\right);
\end{equation}
\begin{equation}
\label{Lor_gauge}
{\cal L}_{gf}=\pi\partial_{\mu}A^{\mu}=\pi\left(\dot A^0+\partial_iA^i\right).
\end{equation}

Here, $\pi$ is a Lagrange multiplier and, at the same time, a momentum conjugate to $A_0$. It is easy to see that, using the effective Lagrangian, the Hamilton function can be constructed according to the usual rule because the terms with derivatives of $A_0$ with respect to time vanish:
\begin{equation}
\label{EPS_Ham}
{\cal H}=\pi\dot A_0+p_i\dot A^i-{\cal L}.
\end{equation}

We have two basic approaches to quantization: the canonical approach relying on the Hamiltonian formalism, and the path integral approach. In the canonical approach, the spacetime topology is restricted by a product of the real line with some three-dimensional manifold. In quantum field theory, the path integral approach was originally used for construction of the $S$-matrix, that implies that particles in initial and final (asymptotic) states are outside the interaction region. In its turn, it means that the path integral is considered under asymptotic boundary conditions which exclude nonphysical degrees of freedom in initial and final states. The asymptotic boundary conditions ensure gauge invariance of the path integral and, therefore, gauge invariance of the whole theory. However, in the case of gravity, the assumption on asymptotic states is valid only in asymptotically flat spacetimes.

Returning to the question about non-trivial topology, we come to the conclusion that both canonical and path integral approaches do not admit an arbitrary topology of spacetime.

Further, we refuse the assumption about asymptotic states, and we shall work in extended phase space that includes, on equal footing, physical, gauge and ghost degrees of freedom. So, we shall come to the formulation of Hamiltonian dynamics in extended phase space, which will be considered in the next Section. It is a prerequisite of quantization, and it explains why the proposed approach has been called the extended phase space approach to quantization of gravity.

\section{The formulation of Hamiltonian dynamics in extended phase space}
This new formulation of Hamiltonian dynamics is based on introducing missing velocities into an effective Lagrangian by means of gauge conditions in a differential form. Thanks to it, the Hamiltonian can be constructed by the same rule as for unconstrained systems.

Varying the effective action, one obtains modified Einstein equations that include additional terms resulting from the gauge fixing and ghost parts of the action. One should add gauge conditions and ghost equations to the modified Einstein equations, so one gets the extended set of Lagrangian equations.

The Hamiltonian set of equations in extended phase space is completely equivalent to the extended set of Lagrangian equations. The equivalence implies that the constraints, gauge conditions and ghost equations are Hamilton equations. Thus, the description of the dynamics appears to be as close as possible to the description of a system without constraints, while the constraints are preserved. They are modified just like other Einstein equations.

The extended phase space approach enables us to solve some problems we face in the Dirac formalism. For example, we know that, in the theory of gravity, different parameterizations of variables are used. The gravitational field can be represented by components of the metric tensor as well as by the Arnowitt -- Deser -- Misner variables. From the viewpoint of the Lagrangian formalism, it is just a change of variables.
\begin{equation}
\label{ADM}
g_{00}=\gamma_{ij}N^iN^j-N^2;\quad
g_{0i}=\gamma_{ij}N^j;\quad
g_{ij}=\gamma_{ij}.
\end{equation}

In theories without constraints, any change of variables in the Lagrangian formalism corresponds to a canonical transformation in the Hamiltonian formalism. However, in the Dirac approach, this change of variables, which touches upon gauge variables, is not canonical.

This change of variables, which is absolutely legal in the Lagrangian formalism, leads to a contradiction from the viewpoint of the Dirac approach. One can check that the Poisson bracket between the lapse function and the momenta conjugate to the spatial component of the metric tensor in not zero \cite{KirKuz,Shest2}:
\begin{equation}
\label{Kir_Kuz}
\left.\{N,\Pi^{ij}\}\right|_{g_{\mu\nu},p^{\lambda\rho}}\ne 0.
\end{equation}

At least, it means that the Dirac Hamiltonian dynamics is not completely equivalent to the original (Lagrangian) formulation of the Einstein theory.

However, when we deal with the effective action and introduce the gauge fixing term, the momenta are modified, which results in correct values of the Poisson brackets.

It has been demonstrated in the full gravitational theory that a transformation of field variables in the Lagrangian formalism touching upon gauge degrees of freedom is a canonical transformation in extended phase space if one chooses a differential form of gauge conditions \cite{Shest2}.

In the Dirac formalism, constraints are generators of transformations in phase space,
\begin{equation}
\label{Dir_gen}
\delta B=\{B,\phi_\alpha\}\varepsilon^\alpha(x).
\end{equation}
For example, in the case of electrodynamics, we have
\begin{equation}
\label{electrod}
\delta A^0=\varepsilon(x),\quad
\delta A^i=\partial^i\xi(x).
\end{equation}

In the theory of gravity, one cannot obtain correct transformations for all degrees of freedom, including gauge ones, using constraints as generators.

In the Batalin -- Fradkin -- Vilkovisky approach the generator (the BRST charge) can be constructed as a series in Grassmannian (ghost) variables with coefficients given by generalized structure functions of constraints’ algebra \cite{Hennaux,Shest3}.
\begin{equation}
\label{Omega_BFV}
\Omega_{BFV}
 =\int d^3x\left(c^\alpha U^{(0)}_\alpha+c^\beta c^\gamma U^{(1)\alpha}_{\gamma\beta}\rho_\alpha+\cdots\right).
\end{equation}

Since the idea of construction of the BRST charge (\ref{Omega_BFV}) is based on the constraints’ algebra, one also cannot get correct transformations for all degrees of freedom, including gauge ones, by means of this generator.

There exist another way to construct the BRST charge making use of global BRST symmetry and the Noether theorem. The BRST charge for the Yang -- Mills fields constructing according to the Noether theorem coincides exactly with the one obtained by the Batalin -- Fradkin -- Vilkovisky prescription.

However, in the case of gravity, the BRST charge constructed according to the method of Batalin, Fradkin and Vilkovisky, differs from the BRST charge constructed by the Noether theorem. The latter (Noether) charge generates the correct transformations for all degrees of freedom, including gauge ones.

This means that the group of transformations generated by gravitational constraints differs from the group of gauge transformations of general relativity in the Lagrangian formalism.

\section{Quantization}
Now I shall turn to quantization. I prefer path integral quantization, since this approach enables us to explore systems without asymptotic states and problems related to gauge invariance. If we consider a gravitating system without asymptotic states, gauge invariance of the theory cannot be proved. It means that the Wheeler -- DeWitt equation, that expresses this gauge invariance, loses its sense. But one can derive a Schr\"odinger equation from the path integral instead, which is believed to maintain a fundamental meaning.

As it is known, for the first time the Schr\"odinger equation was derived from the path integral by Feynman in his seminal paper of 1948 \cite{Feynman}. Then, it was generalized by Cheng \cite{Cheng} for quadratic Lagrangians,
\begin{equation}
\label{Cheng_L}
L(x,\dot x)=\frac12 g_{ij}(x)\dot x^i\dot x^j.
\end{equation}
The Schr\"odinger equation for the Lagrangian (\ref{Cheng_L}) is as following:
\begin{equation}
\label{Cheng_Eq}
i\hbar\frac{\partial\psi}{\partial t}
 =-\frac{\hbar^2}2\frac1{\sqrt{g}}\frac{\partial}{\partial x^i}\left(\sqrt{g}g^{ij}\frac{\partial\psi}{\partial x^j}\right)
  +\frac{\hbar^2}6R\psi.
\end{equation}
Here $g_{ij}(x)$ is a metric of configurational space. $g$ is its determinant, and the quantum correction appears that is proportional to $\hbar^2$ and the curvature $R$ of the configurational space.

We shall consider the effective action for a model with a finite number of degrees of freedom which includes gauge fixing and ghost terms.
\begin{equation}
\label{S_fin}
S=\int dt\left[\frac12 g_{ab}(N,q)\dot q^a\dot q^b-U(N,q)
 +\pi\left(\dot N-\frac{\partial f}{\partial q^a}\dot q^a\right)+N\dot{\bar\theta}\dot\theta\right].
\end{equation}
Again, $g_{ab}$  is a metric of configurational space, $\{q^a\}$ denote physical degrees of freedom and $N$ is a gauge variable, it may be the lapse function or not, depending on the model, $U(N,q)$ is a potential. We introduces the gauge condition $N=f(q)$ in a differential form (like in electrodynamics) and ghost fields $\theta$, $\bar\theta$.

The Hamilton function in extended phase space is:
\begin{eqnarray}
\label{Ham_fin}
H&=&\frac12 g^{ab}p_ap_b+\pi p_a\frac{\partial f}{\partial q_a}
 +\frac12\pi^2\frac{\partial f}{\partial q^a}\frac{\partial f}{\partial q_a}
 -U(N,q)+\frac1N{\cal\bar P}{\cal P}\nonumber\\
&=&\frac12 G^{\alpha\beta}P_{\alpha}P_{\beta}+U(N,q)+\frac1N{\cal\bar P}{\cal P};
\end{eqnarray}
where
\begin{eqnarray}
\label{matrG}
G=\left(\begin{array}{cc}
\displaystyle\frac{\partial f}{\partial q^a}\frac{\partial f}{\partial q_a}
 &\frac{\partial f}{\partial q_a}\\
\displaystyle\frac{\partial f}{\partial q_a}& g^{ab}
\end{array}\right),
\end{eqnarray}
$Q^{\alpha}=\{N,q^a\}$; $P_{\alpha}=\{\pi,p_a\}$.

Now, we need to generalize the Feynman method for constrained systems. As a result, we come to the Schr\"odinger equation,
\begin{equation}
\label{Schr_math}
i\frac{\partial\Psi(N,q,\theta,\bar\theta;t)}{\partial t}=H\Psi(N,q,\theta,\bar\theta;t).
\end{equation}
This equation is a direct mathematical consequence of the path integral with the effective action without asymptotic boundary conditions. I shall refer to it as the mathematical Schr\"odinger equation.

The Hamilton operator $H$ corresponds (up to operator ordering) to the Hamilton function in extended phase space (\ref{Ham_fin}),
\begin{equation}
\label{Ham_op}
H=-\frac1{2M}\frac{\partial}{\partial Q^\alpha}
 \left(MG^{\alpha\beta}\frac{\partial}{\partial Q^{\beta}}\right)+U(N,q)+V[f]
 -\frac1N\frac{\partial}{\partial\theta}\frac{\partial}{\partial\bar\theta},
\end{equation}
Here $M$ is the measure in the path integral, $V[f]$ is a quantum correction which is proportional to $\hbar^2$ and the curvature of configurational space.

The wave function, that is a solution to this equation, defined on extended configurational space. The extended configurational space includes physical and gauge degrees of freedom and ghosts.
\begin{equation}
\label{gen_sol}
\Psi(N,q,\theta,\bar\theta;t)=\int\Psi_k(q,t)\delta(N-f(q)-k)(\bar\theta+i\theta)dk.
\end{equation}
The $\delta$-function fixes the gauge condition (up to a constant $k$). The function $\Psi_k(q,t)$, which depends only on the physical variables, contains information about the physical system.

Substituting this general solution into the mathematical Schr\"odinger equation, we come to the physical Schr\"odinger equation, and the ``physical'' Hamilton operator $H_{(phys)}[f]$ depends on the chosen gauge conditions.
\begin{equation}
\label{Schr_phys}
i\frac{\partial\Psi_k(q,t)}{\partial t}=H_{(phys)}[f]\Psi_k(q,t),
\end{equation}
\begin{equation}
\label{Ham_phys}
H_{(phys)}[f]=\left[-\frac1{2M}\frac{\partial}{\partial q^a}
 \left.\left(Mg^{ab}\frac{\partial}{\partial q^b}\right)+U(N,q)+V[f]\right]\right|_{N=f(q)+k},
\end{equation}
The wave function of the Universe satisfying this equation will describe geometry of the Universe from the point of view of an observer in some fixed reference frame.

\section{Where the extended phase space approach leads}
Now I shall turn to the consequences of this approach. First, I would like to address the question of non-trivial topology which I have already mentioned.

From a theoretical point of view, the path integral enables one to consider spacetimes with non-trivial topology using various coordinates in different regions. Indeed, let us consider a spacetime manifold that includes regions with different gauge conditions. Imagine that the spacetime manifold consists of several regions ${\cal R}_1$, ${\cal R}_2$, ${\cal R}_3,$ \ldots, in each of them different gauge conditions $C_1$, $C_2$, $C_3$, \ldots, being imposed. The regions are separated by boundaries ${\cal S}_1$, ${\cal S}_2$, \ldots. For example, if ${\cal S}_1$ is the boundary between the regions ${\cal R}_1$ and ${\cal R}_2$, one has
\begin{eqnarray}
\label{PI1}
&&\hspace{-1cm}
\int\exp\left(iS\left[g_{\mu\nu}\right]\right)
 \prod\limits_{x\in{\cal M}}M\left[g_{\mu\nu}\right]
 \prod\limits_{\mu,\,\nu}dg_{\mu\nu}(x)\nonumber\\
&=&\int\exp\left(iS_{(eff)}\left[g_{\mu\nu},\,C_1,\,{\cal R}_1\right]\right)
 \prod\limits_{x\in{\cal R}_1}M\left[g_{\mu\nu},\,{\cal R}_1\right]
 \prod\limits_{\mu,\,\nu}dg_{\mu\nu}(x)\nonumber\\
&&\times\exp\left(iS_{(eff)}\left[g_{\mu\nu},\,C_2,\,{\cal R}_2\right]\right)
 \prod\limits_{x\in{\cal R}_2}M\left[g_{\mu\nu},\,{\cal R}_2\right]
 \prod\limits_{\mu,\,\nu}dg_{\mu\nu}(x)
 \prod\limits_{x\in{\cal S}_1}M\left[g_{\mu\nu},\,{\cal S}_1\right]
 \prod\limits_{\mu,\,\nu}dg_{\mu\nu}(x)\times\ldots
\end{eqnarray}

There exists a problem if the boundaries between regions are not spacelike hypersurfaces of equal time. The assumption on an arbitrary topology prevents us from introducing a global time in the whole spacetime, one should rather consider different clocks in every region.

However, now we consider a simple case when the hypersurfaces ${\cal S}_1$, ${\cal S}_2$, \ldots, correspond to some time instants $t_0$, $t_1$, \ldots.

As we can see from (\ref{Ham_phys}), the Hamilton operator in the Schr\"odinger equation in each region will depend on the chosen reference frame (gauge conditions). This Hamilton operator governs time evolution of the gravitational system within the region, and, as in ordinary quantum mechanics, the evolution will be unitary. For example, if $|g_{\mu\nu}^{(0)},{\cal S}_0\rangle$ is an initial state in the region ${\cal R}_1$, the final state in this region is
\begin{equation}
\label{S1}
|g_{\mu\nu}^{(1)},\,{\cal S}_1\rangle=
 \exp\left[-iH_{1(phys)}(t_1-t_0)\right]|g_{\mu\nu}^{(0)},\,{\cal S}_0\rangle.
\end{equation}

Then the evolution continues in the region ${\cal R}_2$, but state vectors in ${\cal R}_2$ belong to another Hilbert space. Therefore, we should expand the final state in ${\cal R}_1$ in a new basis, that means a transformation of this state as
\begin{equation}
\label{projection}
{\cal P}({\cal S}_1,\,t_1)\exp\left[-iH_{1(phys)}(t_1-t_0)\right]|g_{\mu\nu}^{(0)},\,{\cal S}_0\rangle.
\end{equation}
where the operator ${\cal P}({\cal S}_1,\,t_1)$ is a projection operator and it is not unitary in general. In this way, we shall obtain,
\begin{eqnarray}
\label{g3}
|g_{\mu\nu}^{(3)},{\cal S}_3\rangle
&=&\exp\left[-iH_{3(phys)}(t_3-t_2)\right]{\cal P}({\cal S}_2,\,t_2)\nonumber\\
&\times &\exp\left[-iH_{2(phys)}(t_2-t_1)\right]{\cal P}({\cal S}_1,\,t_1)
 \exp\left[-iH_{1(phys)}(t_1-t_0)\right]|g_{\mu\nu}^{(0)},\,{\cal S}_0\rangle.
\end{eqnarray}

From this consideration, we come to the following conclusion:

{\bf At any boundary between the regions with different gauge conditions unitary evolution may be broken down. The operators ${\cal P}({\cal S}_i,\,t_i)$ project the states obtained as a result of unitary evolution in the region ${\cal R}_i$ onto a basis in the Hilbert space in the neighboring region ${\cal R}_{i+1}$.}

Now we shall consider a small variation of the gauge condition. Let $H_{(phys)}[f]$ is a physical Hamilton operator in the region with the gauge condition $N=f(q)+k$ (\ref{Ham_phys}), while $H_{(phys)}[f+\delta f]$ is a physical Hamilton operator in the region with a modified gauge condition $N=f(q)+\delta f(q)+k$,
\begin{equation}
\label{Ham_phys_R2}
H_{(phys)}[f+\delta f]
=\left.\left[-\frac1{2M}\frac{\partial}{\partial q^a}
  \left(Mg^{ab}\frac{\partial}{\partial q^b}\right)
 +U(N,q)+V[f+\delta f]\right]\right|_{N=f(q)+\delta f+k}.
\end{equation}

Bearing in mind that the variation of the gauge condition is small, one can present $H_{(phys)}[f+\delta f]$ in the form:
\begin{equation}
\label{Ham_phys_approx}
H_{(phys)}[f+\delta f]=H_{(phys)}[f]+W[\delta f]+\delta U[\delta f]+V_1[\delta f],
\end{equation}
where $W[\delta f]$ is not Hermitian operator with respect to the basis in the region with a gauge condition $N=f(q)+k$.

Moreover, one can generalize these results for the case of time-dependent gauge conditions. The path integral approach implies that one should approximate the effective action, including the gauge condition, at each small time interval $[t_i,t_{i+1}]$. Let us suppose that changing the gauge condition at the time interval is
\begin{equation}
\label{alter}
\delta f_i(q)=\alpha f_i(q),
\end{equation}
and $\alpha$ is a small parameter, so that
\begin{equation}
\label{step}
N(t)=f(q)+\sum_{i=0}^n\alpha f_i(q)\theta(t-t_i)+k.
\end{equation}
Let us note that, at each time interval, the gauge condition does not depend on time. For example, at the interval $[t_n.t_{n+1}]$ the gauge condition is
\begin{equation}
\label{step_n}
N=f(q)+\sum_{i=0}^{n-1}\alpha f_i(q)+\delta f_n(q)+k.
\end{equation}
and we come to the case considered above. {\bf In the case of time-dependent gauge conditions, it means that at every moment of time we have a Hamilton operator acting in its own ``instantaneous'' Hilbert space. The ``instantaneous'' Hamilton operator is a Hermitian operator at every moment of time, but it is non-Hermitian with respect to the Hilbert space that we had at the previous moment.}

\section{Final remarks}
In conclusion, let us compare the equation (\ref{g3}) with the one describing the evolution of a quantum system according to von Neumann \cite{Neumann},
\begin{equation}
\label{evol}
|\Psi(t_N)\rangle
=U(t_N,t_{N-1}){\cal P}(t_{N-1})U(t_{N-1},t_{N-2})\ldots
 U(t_3,t_2){\cal P}(t_2)U(t_2,t_1){\cal P}(t_1)U(t_1,t_0)|\Psi(t_0)\rangle.
\end{equation}

As is well known, von Neumann wrote that there exist two ways of changing a quantum state of a physical system, namely, unitary evolution and changes as results of measurements over the physical system. In (\ref{evol}), the projection operators ${\cal P}(t_i)$ correspond to measurements made at $t_0$, $t_1$, \ldots, $t_{N-1}$. The analogy between (\ref{g3}) and (\ref{evol}) could be understood if we accept the interpretation of the reference frame as a measuring instrument representing the observer in quantum gravity. At the moments $t_0$, $t_1$, \ldots, $t_{N-1}$, transitions from one reference frame to another take place, and interaction between the measuring instrument (reference frame) and the physical subsystem changes. It makes us go to another Hilbert space.

However, we have already mentioned that, in general, the projection operators are not Hermitian. It leads to the following question:

{\bf Can quantum gravity be the origin of nonunitarity?}

This question remains open and requires further investigations. Many physicists believe that unitarity is an inseparable property of any physical theory that cannot be broken down.

On the other hand, in the framework of unitary evolution, it is not possible to describe irreversible processes we face all around. When one needs to describe such processes, one has to introduce some non-unitary operators artificially, so to say, ``by hands''. In contrast, in the extended phase space approach to quantization of gravity, the emergence of projection operators follows from the logical development of the accepted prerequisites.

It is important to remember that all the conclusions above are consequences of the assumption on a non-trivial topology and the absence of asymptotic states.

These conclusions cannot be obtained in approaches based on the Wheeler -- DeWitt equation or using the assumption on asymptotic states.

The extended phase space approach leads to qualitatively new results which were outlined briefly in this talk.

Now, some words about the perspectives.

The next step in our investigation is to compare the extended phase space approach with other approaches to quantization of gravity. For example, in loop quantum gravity, the so called Ashtekar's variables are used instead of components of the metric tensor. Using these variables, one can explore cosmological models. It is planned to compare the results of loop quantum cosmology with those obtained in the extended phase space approach.

A very important question, in my opinion, is the question about the role of gauge degrees of freedom. Dirac considered them as redundant, but, especially in the theory of gravity, they have a clear geometrical meaning. If we drop them out of the theory, we destroy spacetime continuum that was a great achievement of the relativistic theory. The work with the effective action tell us that ghost fields are also not just auxiliary variables. They are not equal identically to zero. In non-gravitational gauge theories they appeared in inner loops of Feynman diagrams, but in gravity, where there may not be asymptotic states, they may play some additional role. We do not know now what this role is.

In the mathematical Schr\"odinger equation, the wave function is defined on extended configurational space, its coordinates are physical, gauge and ghost degrees of freedom. Ghost degrees of freedom give a contribution to curvature of configurational space, which, as I mentioned, is included to quantum correction to the Schr\"odinger equation. In a fact, extended configurational space is a superspace in the sense that it contains anticommuting coordinates. But it is not a superspace of supersymmetric theories. We have questions: Is it possible to reformulate supersymmetric theories in a way to be compatible with the extended phase space approach? Could it reveal us something new about the role of non-physical degrees of freedom?

At last, we would like to have theoretical predictions that could be verified by observational data. The hope is related with quantum gravitational corrections to the Wheeler -- DeWitt equation, or, in our case, to the Schr\"odinger equation. Unfortunately, the observational data is not exact enough to be compared with the theoretical predictions made in the framework of various approaches to quantization of gravity. Nevertheless, our technical capabilities are quickly developing, as well as our understanding of physical laws.

Let me finish with the quote of John Tyndall \cite{Tyndall}, a British physicist of XIXth century.
\begin{quote}
``\ldots Believing, as I do, in the continuity of nature, I cannot stop abruptly where our microscopes cease to be of use. Here the vision of the mind authoritatively supplements the vision of the eye. By a necessity engendered and justified by science I cross the boundary of the experimental evidence\ldots''
\end{quote}

\section*{Acknowledgements}
I am grateful to the Organizers for the opportunity to present the extended phase space approach to quantization of gravity at this workshop.

\end{document}